# Towards a Cost-Benefit Analysis of Additive Manufacturing as a Service

Igor Ivkić[2,3] [a], Tobias Buhmann[3], Burkhard List[1] and Clemens Gnauer[4]
[1]*b&mi, Wiesmath, AT*
[2]*Lancaster University, Lancaster, UK*
[3]*University of Applied Sciences /* [4]*Forschung Burgenland, Eisenstadt, AT*
*i.ivkic@lancaster.ac.uk, 2210781003@fh-burgenland.at, b@3d60.works, clemens.gnauer@forschung-burgenland.at*

Abstract: The landscape of traditional industrial manufacturing is undergoing a pivotal shift from resource-intensive production and long supply chains to more sustainable and regionally focused economies. In this evolving scenario, the move towards local, on-demand manufacturing is emerging as a remedy to the environmentally damaging practice of mass-producing products in distant countries and then transporting them over long distances to customers. This paradigm shift significantly empowers customers, giving them greater control over the manufacturing process by enabling on-demand production and favouring local production sites over traditional mass production and extensive shipping practices. In this position paper we propose a cloud-native Manufacturing as a Service (MaaS) platform that integrates advances in three-dimensional (3D) printing technology into a responsive and eco-conscious manufacturing ecosystem. In this context, we propose a high-level architectural design for a cloud-based MaaS platform that connects web shops of local stores with small and medium-sized enterprises (SMEs) operating 3D printers. Furthermore, we outline an experimental design, including a cost-benefit analysis, to empirically evaluate the operational effectiveness and economic feasibility in a cloud-based additive manufacturing ecosystem. The proposed cloud-based MaaS platform enables on-demand additive manufacturing and opens up a profit sharing opportunity between different stakeholders.

## 1 INTRODUCTION

Many industrial manufacturing sectors are characterised by resource-intensive production processes, long global supply chains, and a tendency to overproduce relative to actual demand (Westkämper et al., 2016). Product development and sale often takes place in the developed world, while mass production is carried out in countries with low labour costs. Consequently, this leads to significant transport distances for products before they reach the end consumer. Such industrial production is neither environmentally friendly nor sustainable, and it fails to contribute to the promotion of local economies.

The COVID-19 pandemic has further highlighted the reliance on production facilities in emerging and developing countries (Mugurusi and de Boer, 2013). This issue becomes particularly evident when established supply chains are disrupted, as exemplified by the Suez Canal blockade (Chopra and Meindl, 2007; Lee and Wong, 2021). Additionally, since the COVID-19 pandemic, there has been a noticeable shift in society, with customers increasingly favouring regionally and sustainably produced products (Schwilling et al., 2021).

In order to meet the new trend of regional and sustainable purchasing in manufacturing, without resorting to offshore production and long logistics chains, a new approach is required in which small and medium-sized enterprises (SMEs) play a central role. From the customer's perspective, products should be produced in their immediate environment and ***on demand***, with a certain degree of customization (***customised products***) (Lu and Xu, 2019). Similar to cloud computing service models (Wong and Hernandez, 2012), a Manufacturing as a Service (MaaS) approach (Lu and Xu, 2019) could offer customers the opportunity to purchase their own customised products and have them produced locally, utilizing a ***pay-per-use*** or ***pay-as-you-go*** billing model.

Advances in additive manufacturing technology (Wong and Hernandez, 2012) have enabled the local and cost-effective production of a variety of products using three-dimensional (3D) printing technology (Shahrubudin et al., 2019; Mai et al., 2016). A 3D object is first designed on a computer and then sent to a 3D printer, where it is printed layer by layer. Over time, the 3D object designed on the computer becomes a physical product. With the aid of 3D printing technology, the MaaS approach could soon be realised, enabling SMEs to produce (or 3D print) cus-

---
[a]  https://orcid.org/0000-0003-3037-7813

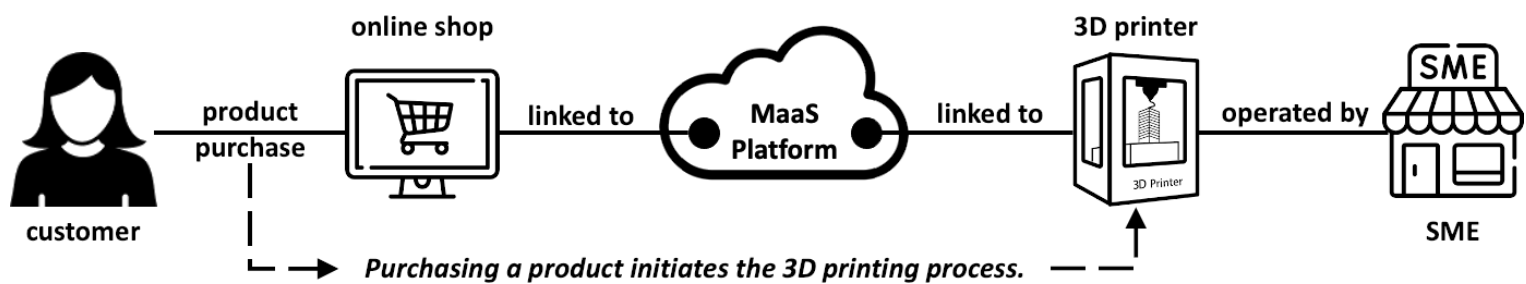
Figure 1: Overall Use Case for an Additive MaaS Platform using 3D Printers.

tomised products *on demand* in close proximity to the customer. This would not only strengthen local economies but also contribute significantly to reducing dependence on offshore production and the associated long supply chains.

Figure 1 shows how a customer's online purchase triggers the 3D printing process, linking the online shop to a cloud-based MaaS platform and subsequently to the 3D printer managed by an SME. This highlights the role of the customer as both the initiator of the *on demand* production process and a proponent of regional, sustainable production.

In this position paper, we present an experimental design to evaluate the feasibility and value of creating a cloud-based additive MaaS platform that uses 3D printers. In this regard, we first introduce the concept of integrating 3D printers with a cloud platform to enable regional and more sustainable production. Building on this, we present the design of an experimental study that uses different metrics to perform a cost-benefit analysis of various 3D printers. This position paper makes a twofold contribution towards additive MaaS using 3D printers:

- Firstly, we introduce a high-level architectural design that enables additive MaaS using 3D printers.
- Secondly, we present the design of a cost-benefit analysis using different metrics and evaluating various 3D printers.

The reminder of the position paper is organised as follows: Section 2 summarises the related work in the field. Next, in Section 3, we describe our experimental design where we introduce an experimental testbed and explain how it could be used in a cost-benefit analysis. Furthermore, we give an overview of potential stakeholders in an additive MaaS ecosystem including a profit sharing approach. Finally, in Section 4 we conclude our work and give an outline of future work in the field.

## 2 RELATED WORK

The integration of additive manufacturing, particularly through a MaaS approach, represents a paradigm shift in industrial production, promising to address the inefficiencies of traditional manufacturing. The following summary gives an overview of existing research, highlighting the advancements and implications of various MaaS technologies and strategies.

### 2.1 Cloud-Based MaaS Approaches

In the field of cloud-based MaaS, several studies have provided contributions that collectively underscore a shift toward more responsive, interconnected, and efficient manufacturing ecosystems, leveraging cloud capabilities to meet the needs of modern production and consumer demand. Chiappa et al. (2023) introduced an open-source project for cloud manufacturing interoperability, highlighting the importance of digital twins, scheduling, and service composition. Thames and Schaefer (2016) proposed a Software-Defined Cloud Manufacturing (SDCM) architecture that enhances flexibility and adaptability in manufacturing systems. Their proposed architecture aligns with the Industry 4.0-paradigm, focusing on flexible hardware systems modifiable at the software level.

Xu (2012) explored the transition from traditional cloud computing to cloud-based manufacturing in the context of Industry 4.0. He introduced concepts of cloud-based manufacturing and outlined its potential to transform traditional manufacturing by utilizing distributed resources and centralised management. Vedeshin et al. (2019) presented a three-layer architecture for a cloud-based manufacturing operating system (OS) that addresses interoperability, scalability, and accessibility and integrates various stages of manufacturing.

Lu and Xu (2019) present a system that integrates cloud-based production equipment with big data analytics for *on demand* manufacturing. This approach addresses challenges like real-time monitoring, and enhances production efficiency and customisation.

## 2.2 3D Printing Technology

Shahrubudin et al. (2019) provide a comprehensive overview of 3D printing technology, covering its types, materials, and applications across industries. They highlight the technology's versatility and its transformative impact on manufacturing processes. Panda et al. (2023) explore 3D printing's role in product development across industries. They detail the stages of development using 3D printing and its advantages and limitations in sectors like healthcare and aerospace. Dhir et al. (2023) analyse how firm size affects the adoption of 3D printing technology by identifying facilitators and inhibitors, providing insightss into the differential impacts on various-sized firms.

## 2.3 Other Related Work

Pahwa and Starly (2021) propose using Deep Reinforcement Learning (DRL) for decision-making in MaaS marketplaces. They demonstrate DRL's effectiveness in improving operational efficiency and profitability over traditional decision-making methods. Bulut et al. (2021) discuss the shift towards MaaS, highlighting its transformative impact on SMEs. The study identifies four key roles in prototype manufacturing and emphasises the need for SMEs to adapt their strategies to thrive in this evolving environment. Chaudhuri et al. (2021) address the gap in MaaS literature regarding pricing strategies. Using game-theoretic models, they propose optimal pricing strategies for MaaS platforms and suppliers. Jagoda et al. (2020), Ranking et al. (2014), and Fiske et al. (2018) explore the feasibility, efficiency, and environmental implications of 3D-printing in military settings.

## 2.4 Summary and Discussion

Building on the identified related work, this position paper contributes to the field by not only proposing a cloud-based MaaS platform that enables local and *on demand* production, but also by detailing an experimental setup to analyse the practicality and value of using different 3D printers. This setup aims to evaluate seven key metrics that will inform the cost-benefit analysis critical to the successful adoption of 3D printing technology by SMEs. In doing so, we aim to bridge the gap between theoretical frameworks and practical, scalable solutions for regional manufacturing. Compared to other related work, the contribution of this position paper is an empirical cost-benefit perspective of cloud-based MaaS specifically tailored for additive manufacturing by proposing an experimental study including different metrics and 3D printers.

## 3 Intended Experimental Design

In this section, we outline an experimental design aimed at conducting a comprehensive cost-benefit analysis to evaluate the overall economic efficiency of our proposed cloud-based MaaS approach. First, we introduce an experimental testbed and describe how it bridges the gap between customers and SME 3D printing production sites through an MaaS platform. We then propose a set of metrics and explain how they can be used to effectively evaluate the economic efficiency of our MaaS approach in a cost-benefit analysis. Additionally, we explore the diverse ecosystem of potential stakeholders in additive MaaS including a profit sharing model. Finally, we provide an outline of future work to carry out the presented experiment.

### 3.1 Experimental Testbed

In a traditional supply chain, raw materials are procured and transformed into finished goods through a sequential process of manufacturing, warehousing, and distribution to retailers, before finally reaching the customer. In contrast, and as shown in Figure 1, the manufacturing process of the MaaS approach is initially triggered after a customer has purchased a product. In other words, the product is not mass-produced and then shipped to a local store where the customers can buy it. Instead, the paying customer initiates the manufacturing process, which starts *on demand* after the product has been purchased.

Once the payment process is complete, the customer's order is sent to a MaaS platform for further processing. This platform plays a central role in connecting various web shops and/or local stores are connected with the SMEs that manufacture (or 3D print) the purchased products. More specifically, the MaaS platform provides API Gateways to connect web shops and Cloud Gateways to connect the SME 3D printing production sites for end-to-end additive MaaS. To implement the overall use case shown in Figure 1, the experimental testbed is divided into the following zones:

- **Zone 1 – Point of Purchase**
  This is where the customers purchase their products and initiate the *on demand* MaaS production process.

- **Zone 2 – Cloud**
  Provides a cloud-native MaaS platform that includes all the gateways to bridge the gap between web shops (**Zone 1**) and the SMEs (**Zone 3**).

- **Zone 3 – SME 3D Printing Production Site**
  Includes all SMEs offering their 3D printing capabilities for additive MaaS.

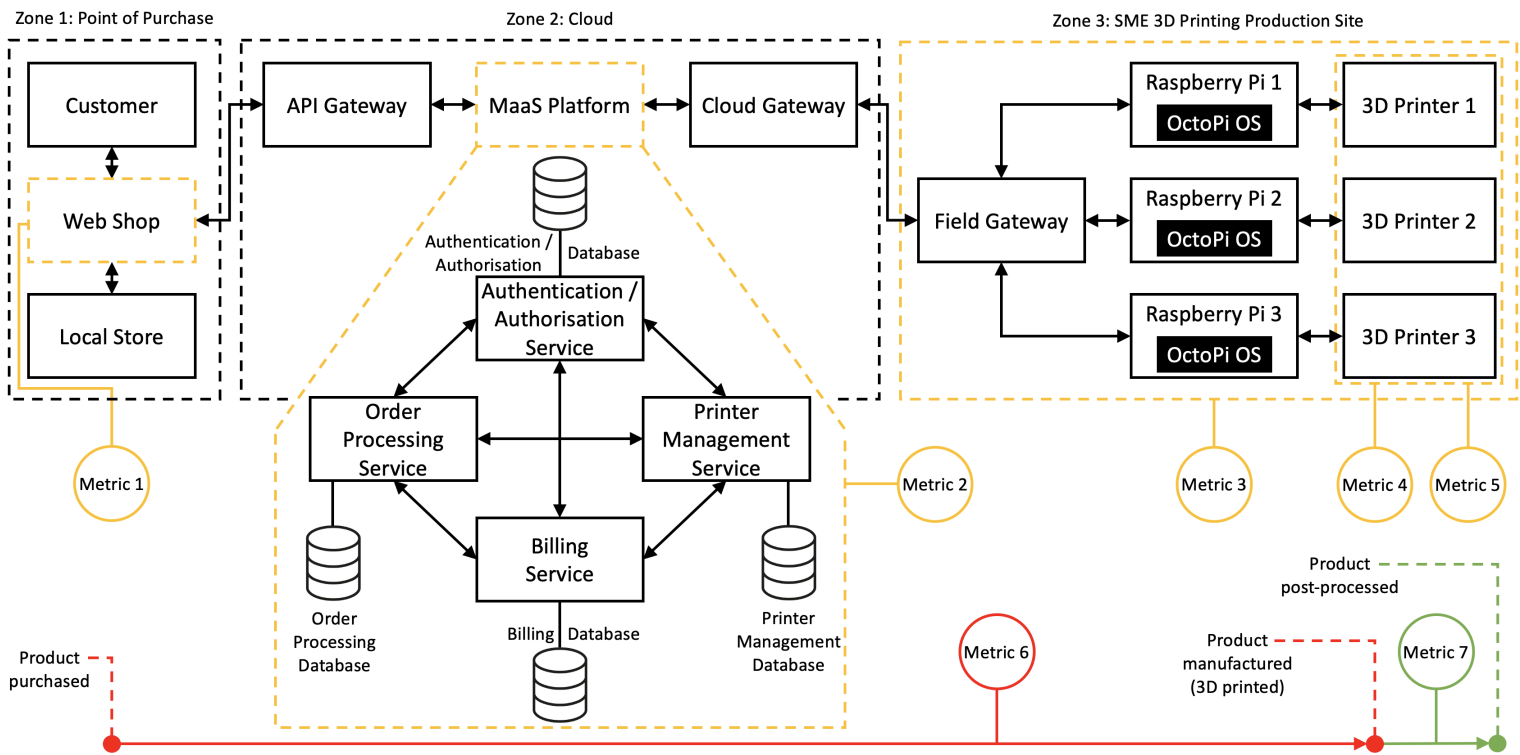

Figure 2: Setup of the Experimental Testbed for Cost-Benefit Analysis of different 3D Printers.

As shown in Figure 2, the experimental testbed is divided into three zones with different tasks and responsibilities. The core component of the testbed is a cloud-based MaaS platform that provides the necessary gateways for connecting the web shops (**Zone 1**) over the cloud (**Zone 2**) with the SME's (**Zone 3**). The platform is based on a cloud-native architecture consisting of four services where each has its own database to ensure loose coupling and service autonomy. This could involve Structured Query Language (SQL) databases for transactional data and Not Only SQL (NoSQL) databases for unstructured data.

The **Order Processing Service** acts as the backbone of the commerce operation, facilitating a seamless transition from customer orders to product manufacturing. It initiates its process once a customer places an order through the web shop. This service is responsible for capturing order details, including product specifications and quantities. It ensures that orders are accurately tracked throughout their lifecycle, providing customers with real-time updates on order and production status. Additionally, it coordinates with the **Printer Management Service** to initiate the manufacturing process, ensuring that the order fulfilment phase begins promptly.

The **Billing Service** is critical for managing the financial transactions within the ecosystem by handling the calculation of costs based on the product specifications. Depending on the purchased product and its complexity additively manufactured the price for the product might vary. Additionally, it generates invoices for customers and SMEs and processes payments supporting various methods. This service plays a pivotal role in ensuring financial integrity and profit sharing between all stakeholders (as elaborated in more detail in Section 3.3).

The **Printer Management Service** maintains a comprehensive registry of all the 3D printers available within the SME network, including their capabilities, configurations, and current operational statuses (available, in use, undergoing maintenance). It dynamically assigns incoming orders to printers based on their location, availability, material requirements, and production capacity. By managing the printers effectively, this service ensures that customer orders are produced accurately and on time, directly influencing the overall throughput and reliability of the service.

The **Authentication / Authorisation Service** is responsible for verifying the identity of all participants in the ecosystem including all user profiles and the 3D printers (devices) themselves. It employs mechanisms to confirm identities and to determine what entity are allowed to do within the MaaS ecosystem. This might include access to specific data, the ability to initiate or cancel orders, and permissions to configure printer settings. This service is crucial for protecting against unauthorised access and ensuring that entities can only perform actions appropriate to their roles, safeguarding the integrity and confidentiality of the ecosystem.

Besides the MaaS platform, **Zone 2** also provides an API Gateway and a Cloud Gateway to connect the web shops from **Zone 1** and the SMEs from **Zone 3**. After a product has been purchased by a customer, the order is sent from the web shop via API Gateway to the MaaS Platform. From there the order is forwarded over the Cloud Gateway to a Field Gateway of a local SME where it is then distributed to one or many 3D printers. For our experimental testbed we have selected three different types of printers from different vendors to increase the validity of the cost-benefit analysis results. In addition, one of the key technical requirements was that all 3D printers should be connected via a serial port to a Raspberry Pi connected to a local network environment (to simulate an SME production site) and receive commands or orders via the Internet. This ensures that a purchase from a web shop actually leads to production by an SME.

In our experimental testbed, each 3D printer (**3D Printer 1–3**) is connected to a corresponding Raspberry Pi (**Raspberry Pi 1–3**), each running the OctoPi OS. This OS provides the so-called *OctoPrint* Representational State Transfer (REST) Application Programming Interface (API) for controlling consumer 3D printers. In addition, each Raspberry Pi is connected to a Field Gateway, which is connected to the cloud-based MaaS platform via a Cloud Gateway. Summarising, the presented experimental testbed from Figure 2 provides all the necessary parts to (1) connect web shops to SMEs via a cloud-based MaaS platform for additive manufacutring *on demand*, and (2) conduct a comprehensive cost-benefit analysis using different metrics and 3D printers (allowing a comparison between different 3D printing devices for more meaningful results).

## 3.2 Cost-Benefit Analysis

In this section, we present an approach to conducting a cost-benefit analysis within the additive manufacturing ecosystem, using the experimental testbed from thee previous section (and as shown in Figure 2). Our methodology involves simulating a complete process, from a *Product purchase* to a *Product manufactured* scenario, using 3D printing technology for a specific product. To enrich our analysis, we include three different 3D printers in our experimental testbed to evaluate and compare their performance in different dimensions. This comparative cost-benefit analysis aims to assess the monetary costs associated with the end-to-end production process of a specific product using a set of seven metrics. These metrics will not only provide a basis for quantifying the cost of 3D printing a specific product, but will also facilitate a comparative analysis of the three different 3D printers. Our aim is to provide a comprehensive view of the cost effectiveness and efficiency of the three 3D printers in a real-world manufacturing environment.

The following metrics have been selected for the proposed cost-benefit analysis:

- **Metric 1 – Operating Costs (Web Shop)**
  refers to the monetary cost of operating a web shop.
- **Metric 2 – Operating Costs (MaaS Platform)**
  refers to the monetary cost of operating the MaaS platform in the cloud.
- **Metric 3 – Energy Consumption**
  refers to the energy cost of operating the SME 3D printing production site including all devices.
- **Metric 4 – Printing Time**
  refers to the time it takes a 3D printer to print a given product.
- **Metric 5 – Material Usage**
  refers to the material cost (amount or quantity of 3D filament used) to print a given product.
- **Metric 6 – Processing Time**
  refers to the lead time it takes to manufacture a purchased product on a 3D printer, starting when the customer purchases the product (Product purchased) and ending when the printer completes the 3D printing process (Product manufactured or 3D printed).
- **Metric 7 – Post-Production Labour Costs**
  refers to the labour costs associated with potential post-production steps (e.g., cleaning, polishing, grinding, painting, etc.).

The above metrics serve as a comprehensive framework for evaluating the cost-effectiveness of 3D printing technologies within our proposed additive MaaS approach. These metrics are can be further categorised into three distinct groups to reflect the multifaceted nature of additive manufacturing. The first category includes metrics that are essential to the operation of online platforms, including the web shops and the cloud-based MaaS platform (**Metric 1, Metric 2**). The second category represents core aspects of the manufacturing process, focusing specifically on energy consumption (**Metric 3**), manufacturing time (**Metric 4**), and material usage (**Metric 5**), providing a granular view of the operational efficiency of devices such as the 3D printers. Finally, the third category, captures the overall processing time for the entire additive manufacturing use case, from product purchase to production completion (**Metric 6**), while also considering post-production labour costs associated with tasks such as cleaning and finishing the product (**Metric7**). This layered approach not only facilitates a detailed cost analysis, but also highlights the operational and environmental considerations that are critical to optimising the additive MaaS processes.

## 3.3 Stakeholders in an Additive MaaS Ecosystem

Building on the proposed experimental testbed and cost-benefit analysis from the previous sections, the following aims to identify potential stakeholders in the additive MaaS ecosystem. In addition, we aim to explore the implications of a profit sharing opportunity between them. This exploration is based on the understanding that the accurate pricing of a product (based on the results of the cost-benefit analysis) is a precursor to the possibility of profit sharing between the identified stakeholders. The following figure shows the identified stakeholders (**Stakeholder 1–4**) and proposes a model for profit sharing:

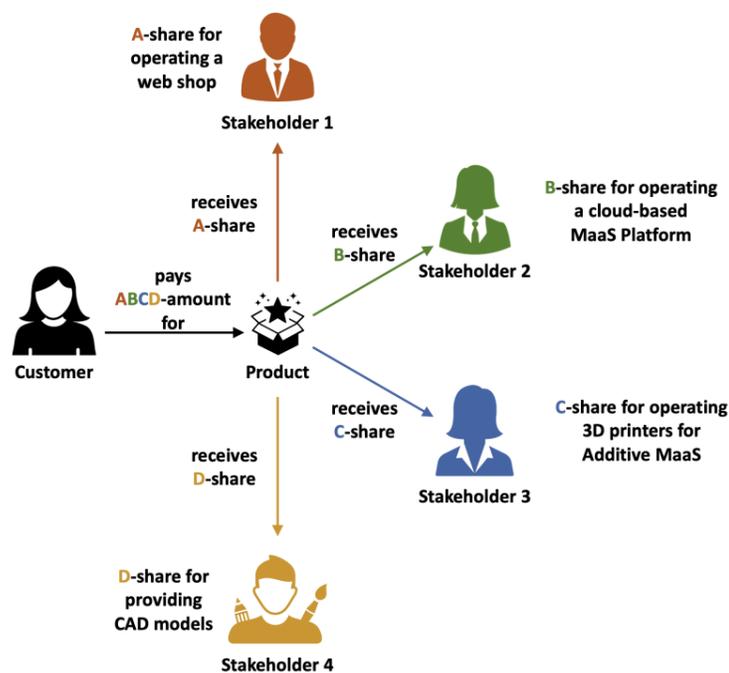

Figure 3: Profit Sharing Opportunity between Stakeholders in an Additive MaaS Ecosystem.

The customer who buys a product does so through a web shop, which is usually operated by a shop owner (**Stakeholder 1**). This means that part of the customer's money belongs to the owner of the web shop used. In order to manufacture the purchased product, the web shop needs to connect to a cloud-based MaaS platform owned and operated by a cloud service provider (**Stakeholder 2**).

This second stakeholder receives a share of the customer's money for providing a MaaS platform to the web shop owner. In return, **Stakeholder 2** forwards each request it receives to the right SME that starts the product manufacturing process using its 3D printers (**Stakeholder 3**). This stakeholder is in turn connected to the cloud-based MaaS platform that enables the SME to receive 3D printing jobs, and receives a share of the initial price paid (by the customer) for the 3D printing service provided.

However, each 3D printer requires a Computer-Aided Design (CAD) model of the product in order to print it. Similar to a developer of mobile phone applications, an additional stakeholder who designs CAD models and offers them to 3D printing SMEs could fill this gap. This stakeholder receives a share of the sales price for providing the CAD models, so that e.g. web shops can offer different designs of a product and SMEs can print them. Summarising, to enable the described profit sharing opportunity between all four stakeholders (**Stakeholder 1–4**), it is necessary to first measure the costs of producing a specific product (cost-benefit analysis), then price the product correctly and define a fair share for each stakeholder.

## 3.4 Future Work

In this section, we have proposed an intended experimental design including an experimental testbed, a cost-benefit analysis, and a profit-sharing approach between identified stakeholders within the additive MaaS ecosystem. In future work, the overall goal is to implement an end-to-end use case from the web shop, through the MaaS platform, all the way to the SME production site (as shown in Figure 2). This will allow the entire process to be simulated, from the customer buying a product online, to the processing of the order in the cloud, to the 3D printing of the product. As a first step, we plan to implement the MaaS platform as a cloud-native application and deploy it on a public cloud platform such as Amazon Web Services (AWS). As a next step, this setup could be deployed on other public cloud platforms (e.g., Microsoft Azure, Google Cloud Platform) to explore the operating costs of different cloud providers. For the SME production site we plan to set up three 3D printers in our local lab and connect them with the MaaS platform.

The intended experimental testbed will allow the simulation of the entire *on demand* additive manufacturing use case. In addition, we plan to use the identified metrics to evaluate the total cost of production associated with the proposed MaaS approach. This evaluation is a first step in determining the pricing of the product under the proposed profit sharing model, ensuring a fair distribution among all stakeholders involved in the MaaS ecosystem.

## 4 CONCLUSIONS

In this position paper, we have discussed the need to rethink traditional manufacturing paradigms by proposing a cloud-native MaaS platform. The MaaS approach emerges as a compelling alternative, promising a shift towards sustainable, regional manufacturing that leverages advances in 3D printing technology. The proposed cloud-based MaaS platform, presented in our experimental testbed and detailed in the cost-benefit analysis, represents a first step towards realising this vision. By integrating additive manufacturing with cloud computing, our approach enables a more responsive and environmentally conscious manufacturing ecosystem.

The proposed experimental design in Section 3 sets the stage for a practical investigation into the feasibility and value of MaaS. The selection of different 3D printers for the experimental testbed and the metrics for evaluating their performance have been established with the intention of providing a thorough understanding of the operational effectiveness and economic feasibility of the additive MaaS approach.

For future work, we plan to conduct the proposed experimental study to empirically validate the theoretical model presented in this paper. This study will test the performance of the selected 3D printers in a cloud-native MaaS setup and by using the defined metrics of operating costs (web shop, MaaS platform), energy consumption, printing time, material usage, processing time, and post-production labour costs. The outcome of this study is expected to provide valuable insights into the cost-effectiveness of the MaaS approach, enabling SMEs to make informed decisions regarding the adoption of 3D printing technologies.

Further research will also explore the scalability of the MaaS model, the integration of a wider range of additive manufacturing technologies, and the development of robust pricing strategies that benefit all stakeholders within the MaaS ecosystem. By expanding the scope of this research, we aim to establish a comprehensive framework for the implementation of MaaS that not only meets the demands of modern consumers for customisation and sustainability but also enhances the competitive edge of SMEs.

## ACKNOWLEDGEMENTS

Research leading to these results has received funding from the Digital Innovation Hub Süd (DIH - Süd) for the innovation project RePro3D funded by DIH SÜD GmbH 2023 - 2024.

## REFERENCES

Bulut, S., Wende, M., Wagner, C., and Anderl, R. (2021). Impact of manufacturing-as-a-service: business model adaption for enterprises. *Procedia CIRP*, 104:1286–1291.

Chaudhuri, A., Datta, P. P., Fernandes, K. J., and Xiong, Y. (2021). Optimal pricing strategies for manufacturing-as-a service platforms to ensure business sustainability. *International Journal of Production Economics*, 234:108065.

Chiappa, S., Videla, E., Viana-Céspedes, V., Piñeyro, P., and Rossit, D. A. (2023). Cloud manufacturing architectures: State-of-art, research challenges and platforms description. *Journal of Industrial Information Integration*, page 100472.

Chopra, S. and Meindl, P. (2007). *Supply chain management. Strategy, planning & operation*. Springer.

Dhir, A., Talwar, S., Islam, N., Alghafes, R., and Badghish, S. (2023). Different strokes for different folks: Comparative analysis of 3d printing in large, medium and small firms. *Technovation*, 125:102792.

Fiske, M., Edmunson, J., Fikes, J., Johnston, M., and Case, M. (2018). 3d printing in space: A new paradigm for america's military. *The Military Engineer*, 110(715):74–77.

Jagoda, J., Diggs-McGee, B., Kreiger, M., and Schuldt, S. (2020). The viability and simplicity of 3d-printed construction: A military case study. *Infrastructures*, 5(4):35.

Lee, J. M.-y. and Wong, E. Y.-c. (2021). Suez canal blockage: an analysis of legal impact, risks and liabilities to the global supply chain. In *MATEC web of conferences*, volume 339, page 01019. EDP Sciences.

Lu, Y. and Xu, X. (2019). Cloud-based manufacturing equipment and big data analytics to enable on-demand manufacturing services. *Robotics and Computer-Integrated Manufacturing*, 57:92–102.

Mai, J., Zhang, L., Tao, F., and Ren, L. (2016). Customized production based on distributed 3d printing services in cloud manufacturing. *The International Journal of Advanced Manufacturing Technology*, 84:71–83.

Mugurusi, G. and de Boer, L. (2013). What follows after the decision to offshore production? a systematic review of the literature. *Strategic Outsourcing: An International Journal*, 6(3):213–257.

Pahwa, D. and Starly, B. (2021). Dynamic matching with deep reinforcement learning for a two-sided manufacturing-as-a-service (maas) marketplace. *Manufacturing Letters*, 29:11–14.

Panda, S. K., Rath, K. C., Mishra, S., and Khang, A. (2023). Revolutionizing product development: The growing importance of 3d printing technology. *Materials Today: Proceedings*.

Rankin, T. M., Giovinco, N. A., Cucher, D. J., Watts, G., Hurwitz, B., and Armstrong, D. G. (2014). Three-dimensional printing surgical instruments: are we there yet? *Journal of Surgical Research*, 189(2):193–197.

Schwilling, T., Schulze, I., Wilts, H., and Du Bois, P. (2021). Circular economy 2021. secondhand in deutschland.

Shahrubudin, N., Lee, T. C., and Ramlan, R. (2019). An overview on 3d printing technology: Technological, materials, and applications. *Procedia Manufacturing*, 35:1286–1296.

Thames, L. and Schaefer, D. (2016). Software-defined cloud manufacturing for industry 4.0. *Procedia cirp*, 52:12–17.

Vedeshin, A., Dogru, J. M. U., Liiv, I., Draheim, D., and Ben Yahia, S. (2019). A digital ecosystem for personal manufacturing: an architecture for cloud-based distributed manufacturing operating systems. In *Proceedings of the 11th International Conference on Management of Digital EcoSystems*, pages 224–228.

Westkämper, E., Löffler, C., Westkämper, E., and Löffler, C. (2016). Visionen und strategische konzepte für das system produktion: Grenzen überwinden mit strategie und technologie. *Strategien der Produktion: Technologien, Konzepte und Wege in die Praxis*, pages 71–237.

Wong, K. V. and Hernandez, A. (2012). A review of additive manufacturing. *International scholarly research notices*, 2012.

Xu, X. (2012). From cloud computing to cloud manufacturing. *Robotics and computer-integrated manufacturing*, 28(1):75–86.